\documentclass{article}
\usepackage{spconf,graphicx, amssymb}
\usepackage{bm}
\usepackage{multirow}
\usepackage{booktabs}
\usepackage{tikz}
\usepackage{amsmath,amsfonts}
\usepackage{pifont}
\newcommand{\cmark}{\ding{51}}%

\usepackage{graphicx}
\usepackage{tabulary}
\usepackage{multirow}
\usepackage{makecell}
\usepackage{boldline}
\usepackage{hhline}

\graphicspath{{./figs/}}
\DeclareGraphicsExtensions{.png}

\usepackage{cite}
\usepackage{bm}
\usepackage[varg]{txfonts}
\usepackage{mathptmx}	
\usepackage{times}
\usepackage{psfrag}

\newcommand{\leftctx}{{\text{L}}}
\newcommand{\rightctx}{{\text{R}}}
\newcommand{\vicinity}{{\text{V}}}

\usepackage{amsmath}

\title{PickNet: Real-Time Channel Selection for Ad Hoc Microphone Arrays}
\name{Takuya Yoshioka, Xiaofei Wang, and Dongmei Wang}
\address{Microsoft, One Microsoft Way, Redmond, WA, USA}

\begin{document}
\ninept
\maketitle
\begin{abstract} 
This paper proposes PickNet, a neural network model for real-time channel selection for an ad hoc microphone array consisting of multiple recording devices like cell phones. Assuming at most one person to be vocally active at each time point, PickNet identifies the device that is spatially closest to the active person for each time frame by using a short spectral patch of just hundreds of milliseconds. The model is applied to every time frame, and the short time frame signals from the selected microphones are concatenated across the frames to produce an output signal. As the personal devices are usually held close to their owners, the output signal is expected to have higher signal-to-noise and direct-to-reverberation ratios on average than the input signals. Since PickNet utilizes only limited acoustic context at each time frame, the system using the proposed model works in real time and is robust to changes in acoustic conditions. Speech recognition-based evaluation was carried out by using real conversational recordings obtained with various smartphones. The proposed model yielded significant gains in word error rate with limited computational cost over systems using a block-online beamformer and a single distant microphone. 
\end{abstract}

%\title{Real-Time Ad Hoc Microphone Arrays in Action}
%\name{Takuya Yoshioka, Xiaofei Wang, Dongmei Wang, William Hinthorn}
%\address{Microsoft, One Microsoft Way, Redmond, WA, USA}
%\email{\{tayoshio,xiaofewa,dowan,wihintho\}@microsoft.com}

%\begin{document}

%\maketitle
% 
%\begin{abstract}

%\end{abstract}
%\noindent\textbf{Index Terms}: Ad hoc microphone array, channel selection, closest microphone identification, meeting transcription
\section{Introduction}
\vspace{-.5em}

The prevalence of personal digital devices equipped with microphones and network connectivity has created a situation where multiple microphones coexist 
in the same physical space. 
In the workplace, people bring smartphones, tablets, and laptops to a meeting room. These devices can be jointly used to capture the meeting audio. 
The acoustic signals obtained at different locations in the room can capture different speakers' voices at different signal-to-noise ratios (SNRs) and direct-to-reverberation ratios (DRRs). Therefore, 
a system leveraging those spatially distributed microphones, or an ad hoc microphone array, would be able to improve various speech processing systems such as automatic speech recognition (ASR).

Our goal is to build a real-time ad hoc microphone array system that works robustly in various real situations. While many approaches were proposed for utilizing the distributed microphones, including both signal processing-based methods~\cite{Tran10,Souden13,Higuchi16,Heymann17b,Boeddeker18} and ASR-oriented approaches~\cite{Wang2018c,Fiscus97}, they do not meet our real-timeness and/or robustness requirements as we will discuss in Section \ref{sec: review}. 
%Many existing methods are based on batch processing, and some methods, especially those for near-field speech extraction, are often vulnerable under realistic conditions. 

%Research of ad hoc microphone arrays 
%dates back to the 2000s~\cite{Raykar05}, yet they have been too fragile for %real applications thus far. % yet it has been far from real applications. %
%The fundamental problem with previously proposed approaches, such as automatic calibration~\cite{Liu07,Liu07b} and blind beamforming~\cite{Araki18,Horiguchi20}, 
%is that they utilize statistics estimated over time by assuming stationarity. 
%However, the device positions, especially those of the cell phones, frequently and rapidly change as people often pick up their phones during conversations, rendering these approaches unstable. 
%They also underperform on conversational audio without speaker diarization~\cite{Horiguchi20}, which is still difficult to perform in real time,  
%because people speak one after another from different locations.  
%This problem can be exacerbated in practice because the audio data from all devices may not arrive on time due to network latency and instability. 
%In addition, most previous methods cannot be executed in real time. 

\begin{figure}[t]
    \centering
    \includegraphics[scale=0.55]{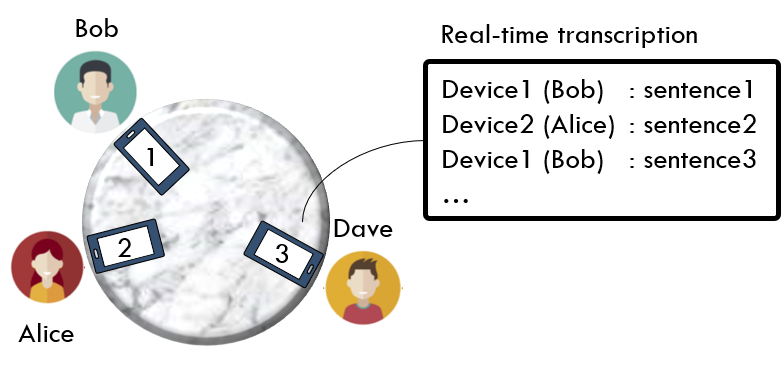}
    \vspace{-1em}
    \caption{Multi-device conversational transcription scenario.}
    \vspace{-1.5em}
    \label{fig: CMDCT}
\end{figure}

In this paper, we adopt a real-time channel selection approach. In conversation scenarios, it is natural that each device is held close to its owner as illustrated in Fig. \ref{fig: CMDCT}. 
Thus, for every short time frame, we identify the microphone that is spatially closest to a currently active speaker with limited latency (64 ms in our experiments). Then, we concatenate the short time frame signals of the estimated closest microphones to produce an output signal. 
By performing the channel selection based on limited acoustic context around each frame, the system can be made robust against the temporal acoustic variations. 
Even when some speakers do not have phones, the method falls back on choosing one of the available microphones from other speakers for the periods where these `non-compliant' speakers are talking.
%---which is not worse than randomly picking one device. 
%For meeting transcription, this approach provides an additional benefit of being able to attribute transcribed words to speakers without voice biometrics by mapping the device IDs to the person names. 

For robust channel selection, we propose a novel neural network model, called PickNet. 
The model leverages spatio-spectro-temporal patterns by interleaving channel-wise convolution layers and cross-channel layers. The cross-channel layer is designed to 
deal with a varying number of input channels in such a way that generates the same outcome regardless of the input channel permutation. 
We evaluated PickNet by using real conversational recordings obtained with multiple cell phones as opposed to the conventional practice of using simulated test samples. 
%which sometimes oversimplifies the complexity of the problem. 
Our PickNet model achieved significantly lower word error rates (WERs) than using a single distant microphone, and its performance came close to or surpassed various non-real-time approaches. 
Note that we use the terms ``device'', ``channel'', and ``microphone'' almost interchangeably in this paper.

\section{Related Work}
\label{sec: review}
\vspace{-.5em}

Technologies that can leverage the spatially distributed microphones have been developed both for speech enhancement and ASR. 
Blind beamforming would be 
the most popular speech enhancement approach as it does not require a priori knowledge about the microphone array's geometry. 
An exemplary blind beamforming method utilizes time-frequency (TF) masks, which are estimated either by clustering~\cite{Tran10,Souden13,Higuchi16} or by using a neural network~\cite{Heymann17b,Boeddeker18}. However, these steps are usually performed with batch or block-online processing, rendering them unusable in real time applications which we are aiming to build. 
Several researchers studied the near-field speech extraction problem~\cite{Taseska15,Furnon21}. 
They assumed each device to be physically close to a different speaker and attempted to extract only the near-field speaker signal from each device. 
Their proposed methods showed promising results when each speaker had one device in their close proximity under certain conditions about the ambient noise. Nonetheless, apparently the performance of these methods deteriorates in some out-of-scope situations, such as the case where each person has two or more near-field devices. These methods are also based on batch processing. 

For ASR, system combination methods such as ROVER~\cite{Fiscus97} can be used to consolidate the ASR outputs from individual devices.
However, it cannot be performed in real time because multiple ASR hypotheses that are generated asynchronously must be aligned at the word sequence level. 
%The high computational cost caused by the need for performing ASR for all channels is also an issue. 
Channel selection is an alternative approach. 
Simple techniques like picking up the audio channel with the maximum energy or SNR were shown to perform poorly~\cite{Wang2018c}. A more effective approach is to make use of phoneme or senone posteriors generated by an acoustic model~\cite{Wang2018c}. While the method of  \cite{Wang2018c} was shown to substantially improve the ASR accuracy, it is based on batch processing and requires acoustic model (AM) evaluation for every input channel. 
%Channel selection is closely related to speech multiplexing because 
%it entails both multiplexing and demultiplexing elements. 
%However, the previous channel selection methods do not necessarily choose the microphone closest to an active speaker. 

%One major drawback of these methods is that they cannot be run in real time because multiple hypotheses need to be aligned although it is possible to reduce the latency to an order of several seconds~\cite{Yoshioka19a}. 

Based on this analysis of the prior work, we set our goal as developing a robust channel selection method that runs in real time with low computational cost. Dealing with overlapped speech is left to future work~\cite{Wang2020,Wang2021}.

\section{Channel Selection Based on Closest Microphone Identification}
\label{sec: methods}
\vspace{-.5em}

We perform channel selection by identifying the microphone closest to the currently active speaker. 
The problem is defined as follows. 
Suppose that we have $M$ spatially distributed recording devices. Let $x_{m,t,f}$ denote the complex-valued short time Fourier transform (STFT) coefficient of the $m$th microphone signal with $t$ and $f$ denoting the time frame and frequency bin indices, respectively\footnote{We assume that each device provides one audio signal. Although modern cell phones often have multiple microphones, the access to the individual microphone channels might not be exposed to app developers.}. 
We estimate the posterior probability, $p_{m,t}$, of the $m$th device being closest to the speaker who is speaking at frame $t$. (Note that we do not consider overlapped speech.)
Then, an enhanced signal, $y_{t,f}$, is computed as the weighted sum of the input STFT coefficients using these posterior probabilities. That is, 
\begin{align}
y_{t,f} = \sum_{m = 0}^{M-1} p_{m,t} x_{m,t,f}. 
\label{eq: enhance}
\end{align}
The estimated STFT signal is converted to a time-domain signal with an inverse STFT. 

In the following description, we omit an index range in sequence notation when the range is evident. Specifically, the ranges of microphone index $m$ and frequency bin index $f$ are assumed to be $(0, \cdots, M-1)$ and $(0, \cdots, F-1)$ unless otherwise indicated where $F$ denotes the frequency bin number. For example, $(a_m)_{0 \leq m \leq M-1}$ is simplified as $(a_m)_m$. 

The set of the channel posterior probabilities for time frame $t$, i.e., $(p_{m,t})_m$, is calculated based on 
the input signals around frame $t$, or $(x_{m,\tau,f})_{m,f,\tau \in \vicinity(t)}$, where $\vicinity(t)$ denotes a vicinity of $t$ defined as $(t - \Delta_{\leftctx}, \cdots, t + \Delta_{\rightctx})$. 
Constants $\Delta_{\leftctx}$ and $\Delta_{\rightctx}$ are the numbers of left (past) and right (future) context frames to be considered. 
$\Delta_{\leftctx}$ should be set to cover the main components of late reverberation, which is essential for identifying the closest microphone. 
$\Delta_{\rightctx}$ must be small enough to keep system's latency at an acceptable level. 
In our system, they were set at 36 and 4, respectively. 
Below, we present two models for the channel posterior probability estimation and our proposed training scheme. 

\subsection{Model without cross-channel layers}
\vspace{-.5em}

\begin{figure*}[t]
    \centering
    \includegraphics[scale=0.5]{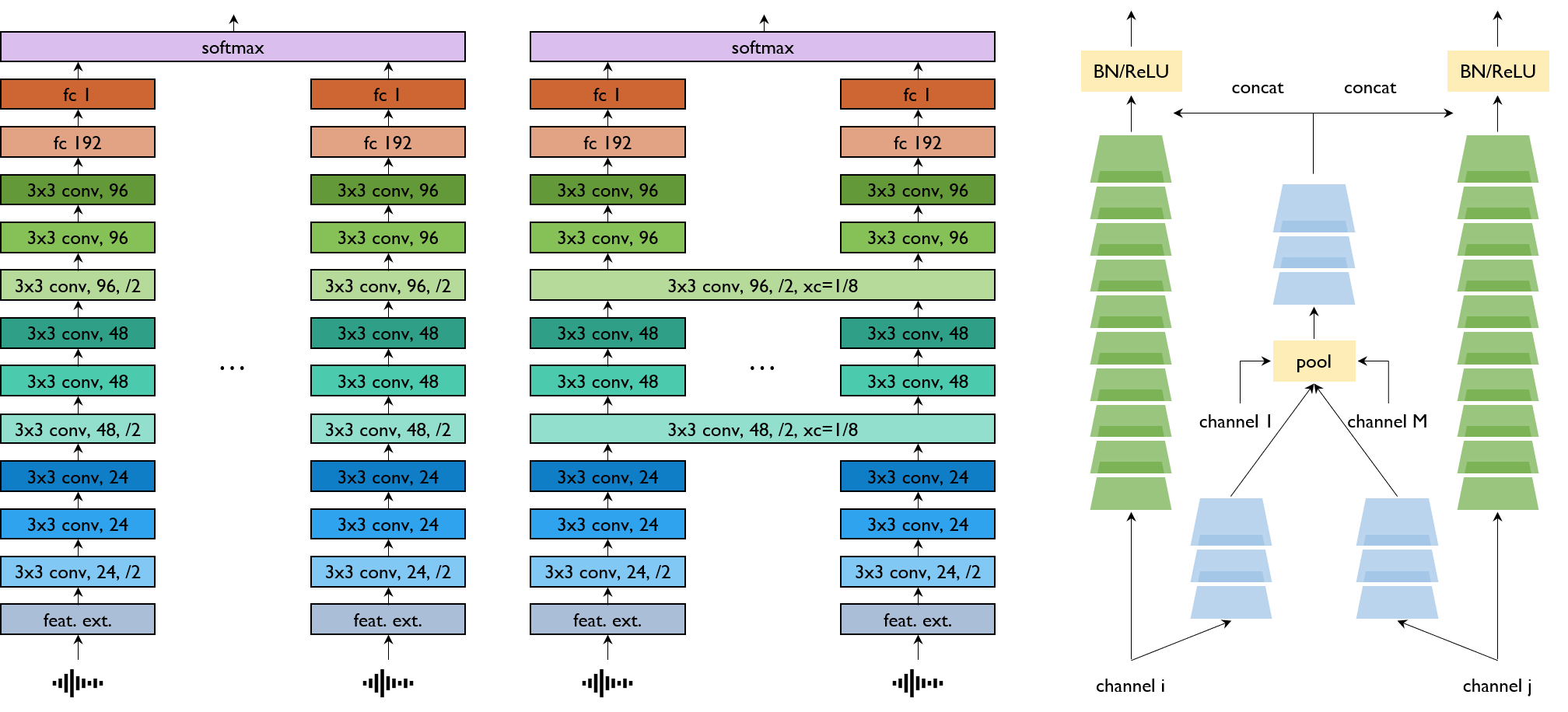}
    \vspace{-1em}
    \caption{Channel selection models. From left to right: (1) model without cross-channel layers; (2) model with cross-channel layers, i.e., PickNet; (3) cross-channel layer. In (1) and (2), layers with the same color share parameters (must be viewed in color). In (2), ``$\textrm{xc}=1/8$'' means one eighth of the convolution kernels are used for cross-channel processing (i.e., kernels producing the feature maps highlighted in light blue in the right diagram). In (3), BN means batch normalization.}
    \label{fig: models}
\end{figure*}

The left diagram of Fig. \ref{fig: models} shows the first modeling approach. Before the last softmax layer, each channel is processed separately with the same model that 
accepts a single-channel signal, $(x_{m, \tau, f})_{f, \tau \in \vicinity(t)}$, as input. 
For each channel, the model generates a scalar which implicitly measures how close the $m$th microphone is to the speaker who is active at time frame $t$.
The $M$ output values are converted to probabilities $(p_{m,t})_m$ with the softmax function. 
We use a convolutional network based on amplitude spectral input.
More specifically, the input to the model is a 2-dimensional tensor comprising the amplitude spectral coefficients of a local TF patch, i.e., $(|x_{m, \tau, f}|)_{f, \tau \in V(t)}$. 
Our model stacks 3x3 convolution layers and fully connected layers to get a scalar output value as shown in the diagram.

%The advantage of this model  is that the neural network computation can be performed independently on each device. Therefore, it requires less computation for a central node (e.g., one particular device or a server).
A drawback of this model is that the closest microphone identification accuracy may be insufficient since each microphone signal is processed independently before softmax.
PickNet overcomes this limitation.

\subsection{PickNet: Model with cross-channel layers}
\vspace{-.5em}

Unlike the model described above, the proposed PickNet model passes around information across the channels by using cross-channel convolution layers.  
This allows the model to compare input channels more effectively by taking account of spatio-specro-temporal patterns.

%This comes at the expense of needing to perform all the computation in one place, such as a remote server. 

The cross-channel layer must deal with a varying number of input channels in such a way that changing the input channel order  results in reordering the outputs while retaining the content of the output for each channel. 
This requirement can be defined as follows. 
Let us represent the mapping realized by the cross-channel layer as $f: (i_0, \cdots, i_{M-1}) \mapsto (o_0, \cdots, o_{M-1})$. 
Now, we consider a permutation, $(q_0, \cdots, q_{M-1})$, of $(0, \cdots, M-1)$. 
Then, $f$ must satisfy $f(i_{q_0}, \cdots, i_{q_{M-1}}) = (o_{q_0}, \cdots, o_{q_{M-1}})$. 

The right diagram of Fig. \ref{fig: models} shows our proposed cross-channel layer, which was inspired by a transform-aggregate-concatenate model~\cite{Luo20,Yoshioka22}. 
The idea is as follows. 
We set aside a subset of the convolution kernels for cross-channel processing. 
The feature maps from these kernels are averaged across all the channels by element-wise mean pooling to generate a set of shared feature maps that capture ``global'' information. 
In each channel, these feature maps are concatenated with the ``local''  feature maps obtained by applying the remaining kernels to the individual channel input. 
As shown in the middle diagram, we convert some of the convolutional layers of the previous model into the cross-channel layers. 
Hence, at the end, we still obtain $M$ scalars and transform them to posterior probabilities with softmax. 
PickNet requires little extra computation compared with the model without the cross-channel layers. 
The cost grows only linearly with respect to the channel number. 

\subsection{Frame subsampling}
\vspace{-.5em}

Reducing a model evaluation cost is crucial for real time processing.  
To this end, we reduces the frequency of calculating the channel posterior probabilities. Specifically, we evaluate the channel selection model only every $N$ frames. 
For the frames where the model is not evaluated, 
the channel posterior probabilities of the preceding frames are used. 
The idea behind this simple modification is that 
the channel to be selected does not change very often because it should change at the same rate as speaker turns in conversations. 
We used $N = 3$ in our system.  
%the efficacy of this scheme is experimentally validated. 

\subsection{Training data and loss function}
\vspace{-.5em}

Training data can be created by simulation from clean speech corpora. 
In our experiments, this was done as follows. 
We used WSJ1 and LibriSpeech~\cite{LibriSpeech} as our clean speech source. 
For each clean speech sample, we determined the dimensions of a virtual shoebox-shaped room for which room impulse responses (RIRs) were generated.
In our simulation, the depth and width values were randomly drawn from a uniform distribution on the interval between 5 and 16 m. 
The height value was uniformly sampled from $[2.5, 4.5]$ m. The room reflection coefficient was randomly determined so that the resultant $\textrm{T}_{60}$ value fell in the range of 0.2 and 0.6 s. 
We assumed that there was one speaker and two microphones in the room, where one microphone should be located near the speaker position. The speaker and near-field microphone positions were randomly determined under the constraint that the speaker-to-microphone distance was between 30 and 70 cm in azimuth and between 10 to 30 cm in elevation. The second microphone position was then determined such that the distance of the two microphones was between 1 and 4 m. Then, the RIR was generated with the image method~\cite{Allen79} and convolved with the clean speech signal to generate a reverberated signal for each microphone position. Hoth noise was added to each microphone signal at a randomly determined signal-to-noise ratio (SNR) between 10 and 20 dB. 
Finally, to account for a transient or impulsive noise sound  captured only by one device (e.g., the noise that is created when the device is forcefully placed on a table), we randomly added a short noise clip taken from the MUSAN corpus~\cite{musan2015} to one of the microphone signals, where the noise clip length was limited to $[0.1, 0.3]$ s. 
%This noise mixing step was found to be critical to obtain a robust model.

As the loss function, 
we used the mean squared error (MSE) between the enhanced signal and the signal of the microphone closest to the speaker, where the MSE is measured for the noiseless condition. 
For each training frame, our model yields posterior probability estimates $(p_m)_m$ from the noisy two-channel input generated as described above. Note that the frame index is omitted for notational simplicity. 
Let $s_{m, f}$ denote the noiseless reverberated signal of the $m$th microphone, which is obtained before the Hoth noise mixing step. 
Also, let the noiseless signal of the near-field microphone denoted as $s_{f}^{\ast}$. 
Then, the loss for this frame is defined as
$\sum_{f} (\sum_{m} p_{m} | s_{m, f} | - | s_{f}^{\ast} | )^2.$
The model parameters are optimized to minimize the average loss computed over the training set. 
The proposed loss function can effectively ignore the silence and short pause segments of the training data by using the signal reconstruction error.

\section{Experimental Results}
\label{sec: experiments}
\vspace{-.5em}

\subsection{System}
\vspace{-.5em}

The proposed channel selection method was evaluated with
a meeting transcription system using multiple cell phones. As illustrated in Fig. \ref{fig: CMDCT}, 
each speaker had one device in his or her vicinity. The system consisted of four steps. 
First, speech signals from individual devices were synchronized with 
correlation matching. The synchronization was repeated every 30 s to account for clock drift errors~\cite{Liu08,Miyabe15}. See \cite[Section~3.1]{Yoshioka19a} for implementation details\footnote{Our informal inspection showed that the synchronization was accurate for most regions. Therefore, we did not investigate its impact on the system performance. 
We used the same synchronization algorithm for all experiments to enable fair comparison.}. 
Secondly, the synchronized signals were fed into the channel selection model, which produced an enhanced signal based on \eqref{eq: enhance} and the posterior probabilities of each device being spatially closest to the active speaker for each time frame. 
Then, streaming ASR was applied to the enhanced signal to generate a word sequence. 

This was done with our internal hybrid ASR system, which was an 
improved version of the one used in \cite{Yoshioka19a}.
Finally, ``device diarization'' (as opposed to speaker diarization) was performed to assign a closest microphone label to each recognized word by using the channel posterior probabilities. 
We used the diarization scheme of \cite{Yoshioka19a} by replacing a d-vector with channel posterior vector $(p_{m, t})_m$.
%Note that the configuration of this experimental system is different from what is being used for the Group Transcribe app. 

Our channel selection module performed STFT by using a 32-ms window with a 16-ms stride. We tested two feature vectors: 257-dimensional amplitude spectra and 80-dimensional log-mel energies. 
The features were mean-normalized along the time axis by using past four seconds of input. 
41 feature frames, comprising a center frame, left 36 frames, and right 4 frames, were stacked to create an input to the channel selection model. 
The processing latency of the channel selection module was the sum of 64 ms (i.e., the 4 right-context frames) and the total of the times for feature extraction and model evaluation.

\subsection{Data and metrics}
\vspace{-.5em}

We collected 12 recordings of real conversations for evaluation. 
Each recording was approximately 20-m long. 
There were four two-speaker conversations, six three-speaker conversations, and two four-speaker conversations. 
The attendees sat around a table and had a natural conversation in English. 
Each person held one cell phone in one hand and put another cell phone on the table mostly within arm's reach. 
This enabled us to examine 
the impact that the device arrangement had on the system performance. 
Various phone models were used, including iOS and Android smartphones of different manufacturers. 
Also, we placed a microphone at the center of the table to capture all participants' voices at an equal distance. 
The conversations were held in various places ranging from small and large conference rooms to an atrium. Therefore, some recordings contained various types of natural background noise. 

Reference human transcriptions were created including speaker labels.
The system outputs were evaluated in terms of WER and word diarization error rate (WDER)~\cite{Shafey2019} by mapping the microphone IDs to the speaker IDs. 
The WDER was used to approximately assess the closest microphone identification performance since we used the real conversational data and thus only word-level ground-truth speaker labels were available. 
Segments including overlapped utterances were excluded from the scoring. 
The computational costs of the proposed models were evaluated with wall clock time measured on a CPU (Intel Xeon CPU E5-2620 v4 2.10GHz) with single threading. Inference was executed with ONNX Runtime for efficient computation.

\begin{table}[t]
\centering
\caption{Evaluation results for proposed models. Column ``Xch'' indicates whether the model has cross-channel layers. Column ``Time'' shows average model evaluation times to process one frame of audio in three-microphone case. The frame interval is 16 ms. 
For WER and WDER, the left and right numbers show the results for hand-held and on-table settings, respectively.}
%\vspace{-.8em}
\label{tab: results}
{\footnotesize
\begin{tabular}{r|cc|ccc|} \cline{2-6}
& Xch & Features & \%WER & \%WDER & Time (ms) \\ \hhline{~=====}

& \multicolumn{2}{c|}{Single distant mic.} & 16.4 & --- & ---\\ \hhline{~=====}
&  & $|\textrm{STFT}|$ & 13.3/14.0 & 3.1/4.9 & 12.1 \\ 
& \cmark (PickNet) & $|\textrm{STFT}|$ & 12.6/14.0 & 2.5/4.9 & 12.1  \\
($\ast$) & \cmark (PickNet) & log-mel & 12.6/13.7 & 2.2/3.0 & 3.97  \\ \hhline{~=====}
 & \multicolumn{2}{c|}{($\ast$) + subsampling} & 12.6/13.8 & 2.3/3.1 & 1.32 \\ \cline{2-6}
\end{tabular}
}
\vspace{-1em}
\end{table}

\subsection{Results}
\vspace{-.5em}

Table \ref{tab: results} shows the evaluation results for different channel selection models and the single distant microphone placed at the table center. 
All models outperformed the single microphone system, demonstrating the usefulness of the channel selection approach. 
We can see that PickNet with log-mel input yielded the best results in terms of both accuracy and processing speed.
We also tested the effect of subsampling by carrying out the model evaluation at an 3-frame interval. 
The result shows that the subsampling significantly reduced the computational cost with little performance degradation. 
Resultant model's cost was small enough to be employed for real usage.

We also conducted an experiment to compare the proposed method 
with three alternative approaches: blind beamforming, AM-based channel selection, and ASR system combination. 
Beamforming was carried out with mask-based beamforming using a model taken from our prior work~\cite{Yoshioka19a}. 
A 1.2-s-long sliding block was used with a stride of 0.4 s and a right context of 0.4 s. 
%Noise spatial covariance matrices were estimated by recursive estimation with a half decay time of 5 s to leverage a longer context. 
As regards AM-based channel selection, we employed M-measure~\cite{Mallidi15,Wang2018c} based on phoneme posteriors, which assesses the degree of the distinctiveness of articulation patterns. 
The phoneme posteriors were computed from the AM output. 
While it was originally used for utterance-wise batch channel selection, we calculated the M-measure values by using a sliding block of 0.8 s with a 0.4-s stride. The input channel with the largest M-measure value was selected for every block position. 
ROVER was used for the system combination. 
Note that all the alternative approaches were not based on real time processing. 
The purpose of this experiment was to clarify where the proposed method stands in terms of efficacy in the broad spectrum of the approaches to the ad hoc microphone arrays.

\begin{table}[t]
\centering
\caption{WER comparison of different multi-channel combination approaches. 
%WDER results for M-measure are shown for reference only because the method is not intended to choose the microphone that is spatially closest to the speaker.
Only PickNet fulfills our real-time processing requirement. This comparison is provided to clarify how the proposed method compares with non-real-time approaches.}
%\vspace{-.8em}
\label{tab: comparison}
%{\footnotesize
%\begin{tabular}{|l|c|cc|} \hline
%Method & Mode & \%WER & \%WDER\\ \hline\hline
%Beamforming~\cite{Boeddeker18} & block online  & 13.6/14.4 & --- \\
%M-measure~\cite{Wang2018c} & block online  & 11.9/13.7 & (4.6/17.2) \\ %ROVER~\cite{Fiscus97} & batch & 14.6/13.7 & --- \\ 
%\hline\hline
%Proposed (PickNet) & real time & 12.6/13.7 & 2.3/3.0 \\ \hline
%\end{tabular}
%}
{\footnotesize
\begin{tabular}{|l|c|c|} \hline
Method & Mode & \%WER \\ \hline\hline
Beamforming~\cite{Boeddeker18} & block online  & 13.6/14.4  \\
M-measure~\cite{Wang2018c} & block online  & 11.9/13.7  \\ ROVER~\cite{Fiscus97} & batch & 14.6/13.7  \\ 
\hline\hline
Proposed (PickNet) & real time & 12.6/13.7 \\ \hline
\end{tabular}
}
\vspace{-1em}
\end{table}

Table \ref{tab: comparison} shows the comparison results. 
We can see that, among the two signal processing-based approaches, 
the proposed method using PickNet always outperformed the blind beamforming method while the former had a much smaller processing latency. This proves the advantage of the channel selection approach in the considered setting using personal devices. 
When the cell phones were held in hands, the proposed method and M-measure significantly outperformed ROVER. This shows the advantage of the channel selection approach over the consensus-based approach under the condition that one input channel is superior to the others. The proposed method yielded a modestly higher WER than M-measure for the hand-held microphone condition. A close inspection revealed that the M-measure-based method produced more stable posterior channel probabilities, which can be attributed to the use of a long right context
and the smoothing effect resulting from the block-based processing. 
Overall, the fact that PickNet was on par with or outperformed the other approaches except for one condition even though it was the only method based on real-time processing demonstrates the effectiveness of the proposed method. 
%The good accuracy of M-measure means that the ineffectiveness of the beamforming approach is attributed to its sensitivity rather than the use of block online processing. 
%A WDER comparison of the proposed method and M-measure also indicates that, in terms of ASR accuracy, there can be one or more microhpones that are as good as or sometimes better than the spatially closest microphones.

%\begin{table}[h]
%\centering
%\caption{Impact of training data configuration with respect to noise %types. All training samples use two microphones as input.}
%\vspace{-1em}
%\label{tab: traindata}
%\begin{tabular}{c|c|c|cc} \hline
%\multicolumn{3}{c|}{Noise seen duing training} & %\multicolumn{2}{c}{Results} \\
%Stationary & Transient & Nonlinear& \%WER & \%SERR  \\ \hline
% $\checkmark$  $\checkmark$  $\checkmark$ & & & 0.0 & 0.0 \\
%$\checkmark$  $\checkmark$ & $\checkmark$   & & 0.0 & 0.0  \\
%$\checkmark$ & $\checkmark$   & $\checkmark$   & 0.0 & 0.0 \\ \hline
%\end{tabular}
%\end{table}

\section{Conclusion}
\label{sec: conclusion}
\vspace{-.5em}

We described a real-time channel selection method for ad hoc microphone arrays based on closest microphone identification. 
Our model, PickNet, can handle a varying number of microphones with the proposed cross-channel convolutional layer, and the performance is invariant to the order with which the microphones are presented to the model. 
An end-to-end multi-device meeting transcription system was built and evaluated by using real conversational recordings, showing the practical utility of the proposed method. 
Based on this work, we built an experimental app called Group Transcribe and released it from the Microsoft Garage project\footnote{We thank the Microsoft Group Transcribe team for the invaluable help they provided throughout this project. The app is currently available for iOS.}.  
The proposed method can potentially be used for a remote-plus-in-person hybrid meeting scenario. The speech quality of the attendees joining from a conference room can be enhanced by using the cell phones or laptops they bring to the room. 
We hope our work can encourage further development of real-time algorithms for ad hoc microphone arrays.

\bibliographystyle{IEEEtran}
\bibliography{my_references}

\begin{thebibliography}{10}
\providecommand{\url}[1]{#1}
\csname url@samestyle\endcsname
\providecommand{\newblock}{\relax}
\providecommand{\bibinfo}[2]{#2}
\providecommand{\BIBentrySTDinterwordspacing}{\spaceskip=0pt\relax}
\providecommand{\BIBentryALTinterwordstretchfactor}{4}
\providecommand{\BIBentryALTinterwordspacing}{\spaceskip=\fontdimen2\font plus
\BIBentryALTinterwordstretchfactor\fontdimen3\font minus
  \fontdimen4\font\relax}
\providecommand{\BIBforeignlanguage}[2]{{%
\expandafter\ifx\csname l@#1\endcsname\relax
\typeout{** WARNING: IEEEtran.bst: No hyphenation pattern has been}%
\typeout{** loaded for the language `#1'. Using the pattern for}%
\typeout{** the default language instead.}%
\else
\language=\csname l@#1\endcsname
\fi
#2}}
\providecommand{\BIBdecl}{\relax}
\BIBdecl

\bibitem{Tran10}
D.~H. {Tran Vu} and R.~{Haeb-Umbach}, ``Blind speech separation employing
  directional statistics in an expectation maximization framework,'' in
  \emph{Proc.\ IEEE ICASSP}, 2010, pp. 241--244.

\bibitem{Souden13}
M.~Souden, S.~Araki, K.~Kinoshita, T.~Nakatani, and H.~Sawada, ``A multichannel
  {MMSE}-based framework for speech source separation and noise reduction,''
  \emph{{IEEE} Trans. Audio, Speech, Language Process.}, vol.~21, no.~9, pp.
  1913--1928, 2013.

\bibitem{Higuchi16}
T.~Higuchi, T.~Yoshioka, N.~Ito, and T.~Nakatani, ``Robust {MVDR} beamforming
  using time-frequency masks for online/offline {ASR} in noise,'' in
  \emph{Proc.\ IEEE ICASSP}, 2016, pp. 5210--5214.

\bibitem{Heymann17b}
J.~Heymann, L.~Drude, and R.~{Haeb-Umbach}, ``A generic neural acoustic
  beamforming architecture for robust multi-channel speech processing,''
  \emph{Comp. Speech, Language}, vol.~46, pp. 374--385, 2017.

\bibitem{Boeddeker18}
C.~Boeddeker, H.~Erdogan, T.~Yoshioka, and R.~{Haeb-Umbach}, ``Exploring
  practical aspects of neural mask-based beamforming for far-field speech
  recognition,'' in \emph{Proc.\ IEEE ICASSP}, 2018, pp. 6697--6701.

\bibitem{Wang2018c}
X.~Wang, R.~Li, and H.~Hermansky, ``Stream attention for distributed
  multi-microphone speech recognition,'' in \emph{Proc. Interspeech}, 2018, pp.
  3033--3037.

\bibitem{Fiscus97}
J.~G. Fiscus, ``A post-processing system to yield reduced word error rates:
  Recognizer output voting error reduction ({ROVER}),'' in \emph{Proc.\ IEEE
  ASRU}, 1997, pp. 347--354.

\bibitem{Taseska15}
M.~Taseska, S.~Markovich-Golan, E.~A.~P. Habets, and S.~Gannot, ``Near-field
  source extraction using speech presence probabilities for ad hoc microphone
  arrays,'' in \emph{Proc. IWAENC}, 2014, pp. 169--173.

\bibitem{Furnon21}
N.~Furnon, R.~Serizel, I.~Illina, and S.~Essid, ``Distributed speech separation
  in spatially unconstrained microphone arrays,'' in \emph{Proc. IEEE ICASSP},
  2021, pp. 4490--4494.

\bibitem{Wang2020}
D.~Wang, Z.~Chen, and T.~Yoshioka, ``Neural speech separation using spatially
  distributed microphones,'' in \emph{Proc. Interspeech}, 2020, pp. 339--343.

\bibitem{Wang2021}
D.~Wang, T.~Yoshioka, Z.~Chen, X.~Wang, T.~Zhou, and Z.~Meng, ``Continuous
  speech separation with ad hoc microphone arrays,'' in \emph{Proc. EUSIPCO},
  2021.

\bibitem{Luo20}
Y.~{Luo}, Z.~{Chen}, N.~{Mesgarani}, and T.~{Yoshioka}, ``End-to-end microphone
  permutation and number invariant multi-channel speech separation,'' in
  \emph{Proc. IEEE ICASSP}, 2020, pp. 6394--6398.

\bibitem{Yoshioka22}
T.~Yoshioka, X.~Wang, D.~Wang, M.~Tang, Z.~Zhu, Z.~Chen, and N.~Kanda,
  ``{V}ar{A}rray: Array-geometry-agnostic continuous speech separation,''
  \emph{arXiv:2110.05745 [eess.AS]}, 2021.

\bibitem{LibriSpeech}
V.~{Panayotov}, G.~{Chen}, D.~{Povey}, and S.~{Khudanpur}, ``{LibriSpeech}: {An
  ASR} corpus based on public domain audio books,'' in \emph{Proc. IEEE
  ICASSP}, 2015, pp. 5206--5210.

\bibitem{Allen79}
J.~B. Allen and D.~A. Berkley, ``Image method for efficiently simulating
  small-room acoustics,'' \emph{J. Acoust. Soc. Am.}, vol.~65, no.~4, pp.
  943--950, 1979.

\bibitem{musan2015}
D.~Snyder, G.~Chen, and D.~Povey, ``{MUSAN}: a music, speech, and noise
  corpus,'' \emph{arXiv:1510.08484 [cs.CD]}, 2015.

\bibitem{Liu08}
Z.~Liu, ``Sound source separation with distributed microphone arrays in the
  presence of clock synchronization errors,'' in \emph{Proc. IWAENC}, 2008.

\bibitem{Miyabe15}
S.~Miyabe, N.~Ono, and S.~Makino, ``Blind compensation of interchannel sampling
  frequency mismatch for ad hoc microphone array based on maximum likelihood
  estimation,'' \emph{Signal Process.}, vol. 107, pp. 185--196, 2015.

\bibitem{Yoshioka19a}
T.~Yoshioka, D.~Dimitriadis, A.~Stolcke, W.~Hinthorn, Z.~Chen, M.~Zeng, and
  X.~Huang, ``Meeting transcription using asynchronous distant microphones,''
  in \emph{Proc. Interspeech}, 2019, pp. 2968--2972.

\bibitem{Shafey2019}
L.~{El Shafey}, H.~Soltau, and I.~Shafran, ``Joint speech recognition and
  speaker diarization via sequence transduction,'' in \emph{Proc. Interspeech},
  2019, pp. 396--400.

\bibitem{Mallidi15}
S.~H. {Mallidi}, T.~{Ogawa}, and H.~{Hermansky}, ``Uncertainty estimation of
  {DNN} classifiers,'' in \emph{Proc. IEEE ASRU}, 2015, pp. 283--288.

\end{thebibliography}

\end{document}